\documentclass[aps,prb,twocolumn]{revtex4-2}

\usepackage{graphicx}
\usepackage[breaklinks=true,colorlinks=true,urlcolor=blue,
citecolor=blue,linkcolor=blue,bookmarks=false]{hyperref}

\begin{document}

	\author{Manabendra Kuiri}
	\thanks{These authors contributed equally}
	\author{Subhadip Das}
	\thanks{These authors contributed equally}
	\author{D V S Muthu}
	\author{Anindya Das}
	\email{anindya@iisc.ac.in}
	\author{A K Sood}
	\email{asood@iisc.ac.in}
	\affiliation{Department of Physics, Indian Institute of Science, Bangalore 560012, India}
	
	\begin{abstract}

		Bulk 1T$^\prime$-MoTe$_2$ shows a structural phase transition from 1T$^\prime$ to Weyl semimetallic (WSM) $ T_{d} $ phase at $\sim$ 240 K. This phase transition and transport properties in the two phases have not been investigated on ultra-thin crystals. Here we report electrical transport, $1/f$ noise and Raman studies in ultra-thin 1T$^\prime$-MoTe$_2$ ($\sim$ 5 to 16 nm thick) field-effect transistors (FETs) devices as a function of temperature. The electrical resistivities for thickness 16 nm and 11 nm show maxima at temperatures 208 K and 178 K, respectively, making a transition from semiconducting to semi-metallic phase, hitherto not observed in bulk samples. Raman frequencies and linewidths for 11nm thick crystal show change around 178 K, attributed to additional contribution to the phonon self-energy due to enhanced electron-phonon interaction in the WSM phase. Further, the resistivity at low-temperature shows an upturn below 20 K along with the maximum in the power spectral density of the low frequency $1/f$ noise. The latter rules out the  metal-insulator transition (MIT) being responsible for the upturn of resistivity below 20 K. The low temperature resistivity follows  $\rho \propto 1/T$, changing to $\rho \propto T$ with increasing temperature supports electron-electron interaction physics at electron-hole symmetric Weyl nodes below 20 K. These observations will pave the way to unravel the properties of WSM state in layered ultra-thin van der Waals materials.
	\end{abstract}

	\title{Thickness dependent transition from 1T$^\prime$ to Weyl semimetal phase in ultrathin MoTe$ _{2} $: Electrical transport, Noise and Raman studies}


	
	\maketitle
	
	\section{Introduction}
	Semimetallic transition metal dichalcogenides (TMDs) have spurred tremendous interest in recent years due to their novel electronic properties and a wide range of potential applications \cite{wang2012electronics}. These materials exhibit exciting phenomena like gate tunable superconductivity \cite{PhysRevB.5.895,sajadi2018gate}, charge density waves \cite{PhysRevB.16.801} and anomalous quantum Hall effect \cite{kang2019nonlinear}. It has been shown that the T$_d$ phase of semimetallic TMDs belong to a class of type-II Weyl semimetals (WSM) \cite{soluyanov2015type, PhysRevLett.117.056805,wang2016gate,qian2014quantum,PhysRevB.86.115133,PhysRevB.88.104412}. A WSM consists of topologically protected linear bands, touching at points referred to as 'Weyl nodes', where the band dispersion is linear in all the three momentum directions  \cite{xu2015experimental,xu2015discovery,lv2015observation,yang2015weyl}. Not only different phases but also tunability of electronic properties of semimetallic TMDs by varying the number of layers have shown thickness-dependent transition temperature of the charge density wave in 1T-TiSe$_2$ and 1T-VSe$_2$  \cite{goli2012charge,yang2014thickness} as well as  magnetic and topological transitions \cite{PhysRevLett.122.107202}. However, the  thickness-dependent electrical transport properties of semimetallic TMDs have not been explored well.

	In this context MoTe$_2$ is a quite promising TMD for the following reasons. MoTe$_2$ crystallizes in normally three stable phases : hexagonal ($\alpha$ or 2H phase), monoclinic ($\beta$ or 1T$^\prime$) and orthorhombic ($\gamma$ or T$_d$ phase) \cite{puotinen1961crystal}. The 2H and 1T$ ^{\prime} $ phases are semiconducting in nature with  bandgap of 1.1 and 0.06eV  respectively\cite{ruppert2014optical,keum2015bandgap}, whereas  the T$ _{d} $ phase is  a type-II WSM \cite{qi2016superconductivity}. Bulk 1T$^\prime$-MoTe$_2$ undergoes a first order structural phase transition from monoclinic 1T$^\prime$ to orthorhombic T$_d$ phase below $\sim$ 240 K \cite{doi:10.1021/acs.chemmater.6b04363,clarke1978low,qi2016superconductivity}. Recent Raman studies \cite{zhang2016raman,chen2016activation,PhysRevB.97.041410} as well as ARPES measurements \cite{huang2016spectroscopic,deng2016experimental,berger2018temperature} confirm the bulk T$_d$ phase of MoTe$_2$ to be a type-II WSM \cite{soluyanov2015type, PhysRevLett.117.056805, PhysRevB.92.161107}. Recent magneto-transport measurements \cite{song2018few} of a few layer 1T$^\prime$-MoTe$_2$ confirm its gapless semimetallic behavior. However, the transition from 1T$^\prime$ to T$_d$ phase and their electrical transport properties, governed by the predicted electron-electron interactions at the Weyl nodes \cite{PhysRevLett.108.046602} have not been investigated on ultra-thin MoTe$_2$ crystals.

	In this work, we report temperature dependent resistivity measurements of few layer 1T$^\prime$-MoTe$_2$ devices with thickness ranging from  5 to 16 nm. We observe a clear signature of 1T$^\prime$ to T$_{d} $ phase transition in 11 nm and 16 nm thick nanocrystals with a transition temperature ($T_{c} $) of 178 K and 208 K, much less than the bulk transition temperature (240 K). For $T>T_c$, semiconducting behavior with $d\rho/dT <0$ is seen which changes to   $d\rho/dT>0$ below $T_{c}$, signifying semimetallic behavior. Such resistivity change is not seen in 1T$^\prime $-MoTe$_{2} $ with thickness of 7 nm and 5 nm, establishing the suppression of transition in crystals with thickness below a critical value. In addition for the 11 nm thick crystal, the resistivity increases below $T \sim 20 K$. In the low temperature regime ($T<20 K$), the resistivity shows $1/T$ behavior, a clear signature of the WSM state as predicted theoretically\cite{PhysRevLett.108.046602}. This feature has not been reported in bulk  \cite{qi2016superconductivity} but seen in 6 nm thick 1T$^\prime$-MoTe$_2$ \cite{song2018few} where the upturn at $\sim $20 K has been attributed to MIT \cite{song2018few}. We have used $1/f$ noise spectroscopy as a function of temperature to look at this suggestion of MIT, as noise spectroscopy has been used recently in VO$ _{2} $ to examine the MIT \cite{topalian2015resistance}. We further note that  $1/f$ noise measurements have been utilized as a sensitive tool to probe structural phase transitions\cite{PhysRevLett.102.025701,barone2016unravelling}, vortex flux dynamics near superconductor-normal phase transition\cite{PhysRevB.76.134515} and electronic phase transition\cite{PhysRevB.83.085302,barone2013electric}. We will show that our $ 1/f $ noise spectra do not agree with this conjecture and propose that it is related to electron-electron interaction physics in the WSM state arising from the current carrying electron-hole states at the Weyl nodes \cite{PhysRevLett.108.046602}. Furthermore, we have carried out Raman scattering experiment as a function of temperature on 11 nm thick MoTe$ _{2} $ device and show that Raman mode frequencies and linewidth exhibit slope change at $T_{c} $ which further supports the transition from 1T$^\prime$ to WSM T$_{d}$ phase observed in our electrical transport measurements.
	
	\begin{figure*}[t!]
		\includegraphics[width=1\textwidth]{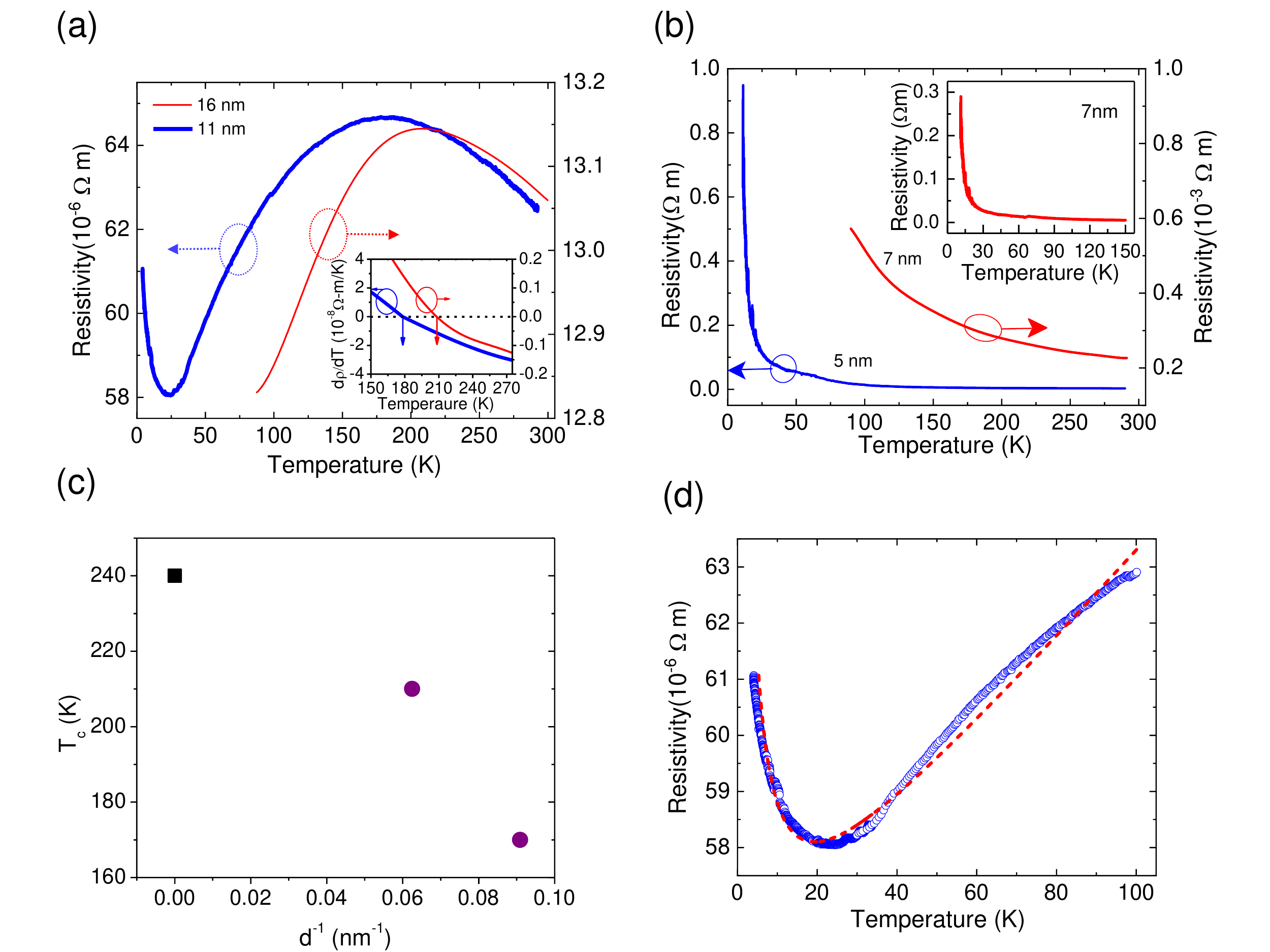}
		\caption{(Color Online) (a) Resistivity ($\rho$) versus temperature (T) for $\sim$ 11 nm and $\sim$ 16 nm nanocrystals. Inset shows $d\rho/dT$ with temperature. $T_c$  is indicated with vertical arrows. (b) Resistivity ($\rho$) versus temperature (T) for thin nanocrystals  $\sim$ 5 nm and $\sim$ 7 nm respectively. Inset shows the resistivity data in the low temperature regime (10 to 150K) of another 7 nm thick device. (c) Critical temperature ($T_c$) as a function of inverse of nanocrystal thickness ($d^{-1}$) (circle points) from Fig. 1(a).  The error bar size is $\sim$ 0.5K. The bulk $T_c\sim$ 240 K (Square dot) has been taken from literature \cite{doi:10.1021/acs.chemmater.6b04363,clarke1978low,qi2016superconductivity}. (d) Resistivity ($\rho$) versus temperature (T) in the low temperature regime (4 K to 100 K) for $\sim$ 11 nm nanocrystal. The red dashed line shows the fit with Eq. (\ref{eqn:fit_rho}).}
		\label{g1}
	\end{figure*}
	
	\section{Experimental details}
	Thin nanocrystals (between 5 and 16 nm) of 1T$^\prime$-MoTe$_2$ were obtained from mechanical exfoliation of bulk single crystals of 1T$^\prime$-MoTe$_2$ onto a piranha cleaned Si/SiO$_2$ substrate with 285 nm SiO$_2$ \cite{kuiri2016enhancing} with immediate  spin-coating with PMMA to prevent degradation \cite{kuiri2015probing}. Electrical contacts were patterned using electron beam lithography followed by thermal deposition of Cr/Au (5 nm/70 nm). The nanocrystals were characterized by atomic force microscopy (AFM) (see Fig. S1 of the supplemental material).

	Resistivity measurements were carried out in a home built dip-stick cryostat with a vacuum $\sim 10^{-6}$ mbar \cite{kuiri2015probing}. Two probe resistance measurements were carried out in a standard lock-in technique, by applying a small ac signal and measuring the current through the lock-in amplifier. The resistance fluctuations were recorded using a NI USB 6210 DAQ card. The details of noise measurement techniques have been discussed extensively in our earlier work \cite{kumar2016tunability}. Raman spectra were recorded using confocal Raman spectrometer (LABRAM HR-800) system using excitation wavelength of 532 nm with low incident power ($ \sim $ 0.5 mW ). The sample temperature was controlled using a liquid nitrogen cooled cryostat (M/s Linkham Scientific Thms350v).
	
	
	
	\section{Results and discussion}

	\subsection{Transport study}

	Fig.~\ref{g1}(a) shows the resistivity ($\rho$) as a function of temperature ($T$) for two devices with thickness of 11 nm and 16 nm. It clearly shows that the resistivity first increases with the decreasing temperature ($d\rho/dT<0$) as expected for the 1T$ ^{\prime} $ semiconducting phase, till a critical temperature $T_c$ of 178 K for 11 nm thick 1T$^\prime$-MoTe$_2$. Below $T_c$, the resistivity decreases ($d\rho/dT>0 $) showing semimetallic behavior until $T\sim 20 K$. A similar trend was also observed for 16 nm thick crystal with $T_c\sim208K$, showing clearly that $T_c$ is suppressed as nanocrystal thickness is reduced. Inset of Fig. \ref{g1}(a) shows derivative of the resistivity  with temperature to determine $T _{c} $.  Our observations are in variance with the reports by He \textit{et al.}, where they observe semimetallic behavior (i.e $d\rho/dT>0$) from 280 K down to 150 K for all sample of thickness 7, 50, 180 nm and 100 $\mu$m \cite{PhysRevB.97.041410}. Secondly, we do not observe hysteretic behavior in resistivity for 11 nm nanocrystal as observed in bulk \cite{qi2016superconductivity,PhysRevB.97.041410} (see Fig. S2(a)). For $T>T_c$, the resistivity shows Arrhenius behavior with activation gap of $\sim 4$ meV. Notably for thin nanocrystals ($\sim$ 5 nm and 7 nm) shown in Fig.~\ref{g1}(b), the resistivity increases monotonically as the temperature is reduced without a change in $ d\rho/dT $. These devices remains semiconducting in the measured temperature range where the resistivity follows Arrhenius equation with activation energy of $\sim$ 3.5 and 3.3 meV for 5 and 7 nm nanocrystals  respectively (Figs. S2(b) and (c)). Thus, only the nanocrystals  with thickness $\geq$ 11 nm show a transition  from 1T$^\prime$ to T$_d$  (WSM) phase. Although we have performed our electrical measurement in two probe configuration, the linear current to voltage relation (see Fig. S2(d)) of our 5nm thick sample at gate voltages ranging from -20V to 20V, suggests negligible  Schottky barrier in our device fabrication process. In Fig.~\ref{g1}(c), we have plotted the transition temperature with the inverse of nanocrystal thickness for our 11 and 16nm samples with the reported $T _{c} $ of the bulk crystal\cite{doi:10.1021/acs.chemmater.6b04363,clarke1978low,qi2016superconductivity}. 
	
	A qualitative understanding of the downward shift of $T _{c} $ is as follows\cite{PhysRevB.97.041410, origin}: In the T$ _{d} $ phase, the Fermi energy lies above the out-of-plane hole bands (marked as $\psi_{1}$ and $\psi_{2}$ in Ref. \onlinecite{origin}). With decreasing thickness, the confinement energy shifts the hole bands to lower energy and hence favors the formation of T$ _{d} $ phase at a lower temperature as compared to bulk 1T$ ^\prime $-MoTe$ _{2} $ (See the Figs. 4(a) and (b) of Ref. \onlinecite{PhysRevB.97.041410}). The reason that T$ _{d} $ phase is not seen up to 10 K for 5nm and 7nm thick nanocrystals  is because of the effect of unintentional hole doping. Gate dependent conductivity measurement (Fig. S2(e))  shows that the 5nm sample has conduction minima around +19 V of gate voltage (indicated by vertical dotted line in Fig. S2(e)) and not 0 V. Therefore, it is evident that at V$ _{bg} $=0V, the device is unintentionally hole doped. The explanation for this unintentional doping is following. It has been reported that rapid thermal annealing of 2H-MoTe$ _{2} $ by oxygen at 250 $ ^{0} $C shows p-type doping due to formation of Mo-O bonds in tellurium vacancy sites \cite{doi:10.1002/adma.201606433}. Similar effect is also observed on this material for prolonged exposure in the atmosphere \cite{6872352}, which can also be expected for 1T$ ^{\prime} $-MoTe$ _{2} $. The effect of unintentional hole doping will be more in the thin samples due to relative increase in surface area and stabilize the 1T$ ^{\prime} $ phase \cite{origin}. Paul \textit{et al.} \cite{paul2019controllable} show that 2 to 5 layers of MoTe$ _{2} $ are in 1T$ ^{\prime} $ phase at room temperature as indicated by the 130cm$ ^{-1} $  Raman mode \cite{zhang2016raman, chen2016activation,PhysRevB.97.041410}. By electron doping these samples using NH$ _{3} $ exposure, the 130cm$ ^{-1} $ mode show splitting, indicative of the T$ _{d} $ phase \cite{zhang2016raman, chen2016activation,PhysRevB.97.041410}. More theory work is needed to quantitatively understand this effect.
	
	\begin{figure*}[t!]
		\includegraphics[width=1\textwidth]{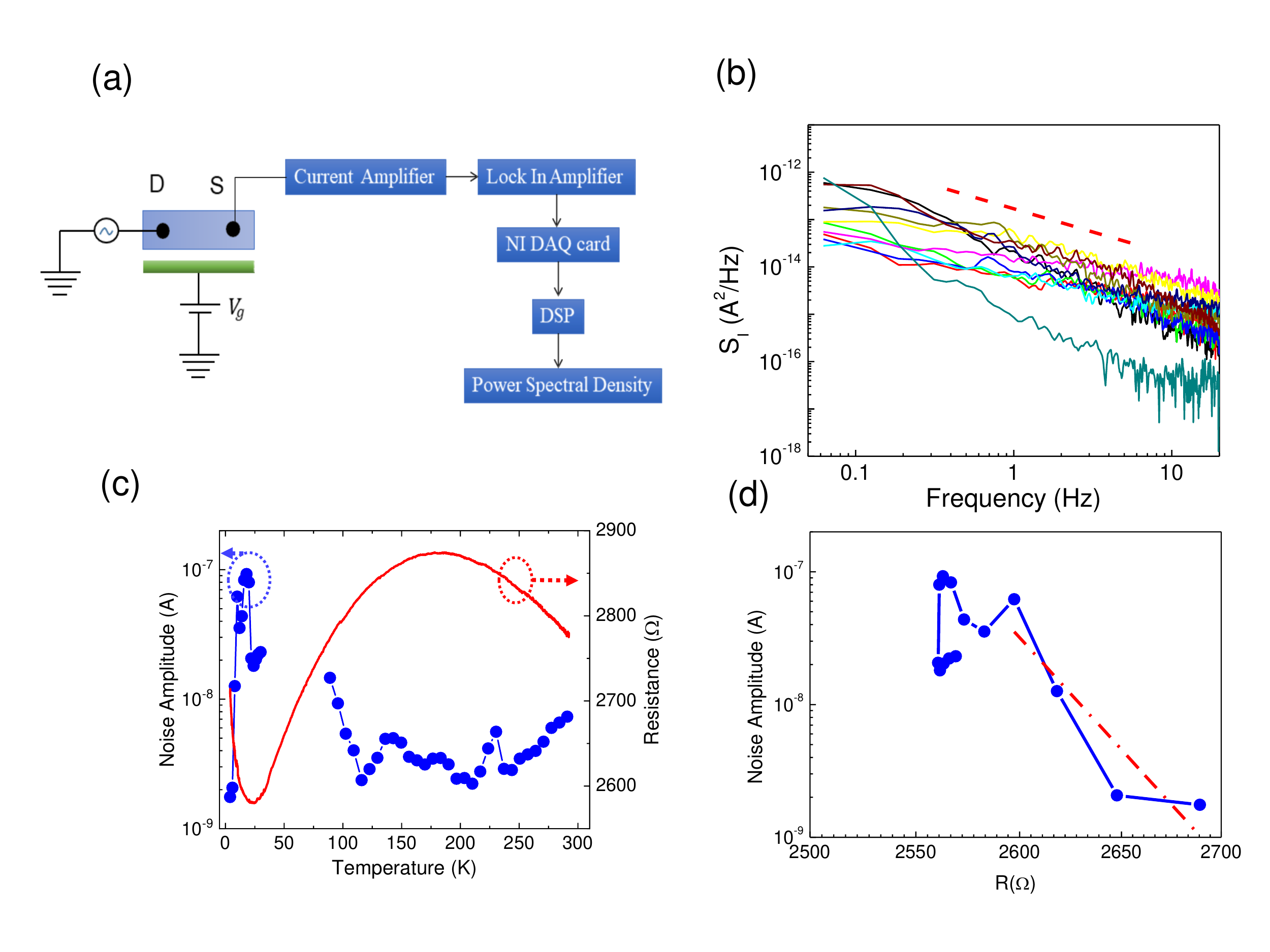}
		\caption{(Color Online) (a) Schematic of $1/f$ noise measurement. (b) Power spectral density for several values of temperatures ($T$). (c) Noise amplitude (A) - blue solid symbols and resistance versus temperature - red line. (d) Noise scaling: Noise amplitude with resistance.}
		\label{g2}
	\end{figure*}
	
	We now discuss the low temperature resistivity (4-100 K)  for the 11 nm nanocrystal (zoomed part of Fig.~\ref{g1}(a)) to focus on the resistivity minimum at $\sim$ 20 K (Fig.~\ref{g1}(d)). Similar upturn in resistivity at temperatures below 10 K was observed in 6 nm 1T$ ^\prime $-MoTe$ _{2}$ crystal \cite{song2018few}, which was attributed to enhanced e-e interactions in two dimension where the resistivity $\rho \propto$ log$_{10}$(T). For 3.6 nm nanocrystal, it was shown that the resistivity increases as the temperature is lowered from ambient to 1.5 K, which was attributed to strong localization effect \cite{song2018few}. It was argued that the lower value of the electron dephasing length ($L_\Phi$) in ultrathin crystals leads to enhanced localization, resulting in MIT at low temperatures \cite{song2018few}. However, our low-temperature upturn resistivity data of 11 nm thick nanocrystal do not scale as log$_{10}$(T) or follow MIT scaling power law as observed in low frequency noise measurements \cite{topalian2015resistance}.  We will come back to it when we discuss the noise measurements in the next section. Regarding the upturn of resistivity below 20 K, we note that the theoretically predicted  resistivity in the WSM state in the clean limit has electron-electron interactions with zero total momentum transfer due to electron-hole symmetry near the Weyl nodes that leads to $1/T$ dependence \cite{PhysRevLett.108.046602}. Guided by this, we fit our resistivity data with the equation
	\begin{equation}
		\rho=\rho_0+\frac{a}{T}+bT
		\label{eqn:fit_rho}
	\end{equation}
	The third term on the right hand side, linear relation of $\rho$ with $T$, is due to electron-phonon interactions, whose nature is determined by the Bloch-Gruneisen temperature ($\Theta_{BG}$) rather than the Debye temperature ($\Theta_{D}$) for smaller Fermi surfaces around the Weyl nodes as observed in graphene \cite{efetovprl2010}.	Fig.~\ref{g1}(d) shows the fit to Eq.(\ref{eqn:fit_rho}), where we observe that for $T<20 K$, $\rho \propto 1/T$ dominates and for $T>20 K$, $\rho \propto T$. We note that addition of a term in resistivity proportional to T$ ^{2} $ arising from electron-electron interaction in Fermi liquid  does not fit the data.

	\subsection{Noise measurements}

	Fig.~\ref{g2}(a) shows the schematic of the $1/f$ noise measurements. Two probe voltage noise was measured using the technique described in our earlier work \cite{kumar2016tunability}. The linear current variation at low bias voltage (Fig. S2(d)) of our 5nm nanocrystal at 77K show negligible Schottky barrier at the interface of channel and electrode and hence rules out the possibility of any  spurious contribution to noise measurements from the device contacts. Fig.~\ref{g2}(b) shows the typical power spectral density (PSD) for several temperatures where the dashed line shows the slope-1. The PSD in our case is $\sim1/f^{\beta}$, where $\beta$ lies between 0.8 to 1.2. Fig.~\ref{g2}(c) shows the comparison of the noise amplitude (\textit{A}) and resistance with temperature. The noise amplitude \textit{A}, changes by two orders of magnitude with temperature, with a maximum at $\sim 20 K$. In Fig.~\ref{g2}(d), we re-plot the Noise amplitude versus resistance (R) to examine the expected scaling seen earlier near MIT \cite{topalian2015resistance}. The noise scaling close to MIT transition is ${S_I}/{I^2} \propto R^x $, where $x$ is the universal exponent \cite{topalian2015resistance}. For MIT, $x \sim 2.6$ \cite{topalian2015resistance}. In comparison, in our case $x\sim100$ which is shown by the dotted line in Fig. \ref{g2}(d). Thus the scaling of noise with resistance near 20K transition is not compatible with MIT. We hope that our experiments will motivate theoretical understanding of noise fluctuations in Weyl semimetals. 
	
	\subsection{Raman spectroscopy}
	
	\begin{figure*}[t!]
		\includegraphics[width=0.7\textwidth]{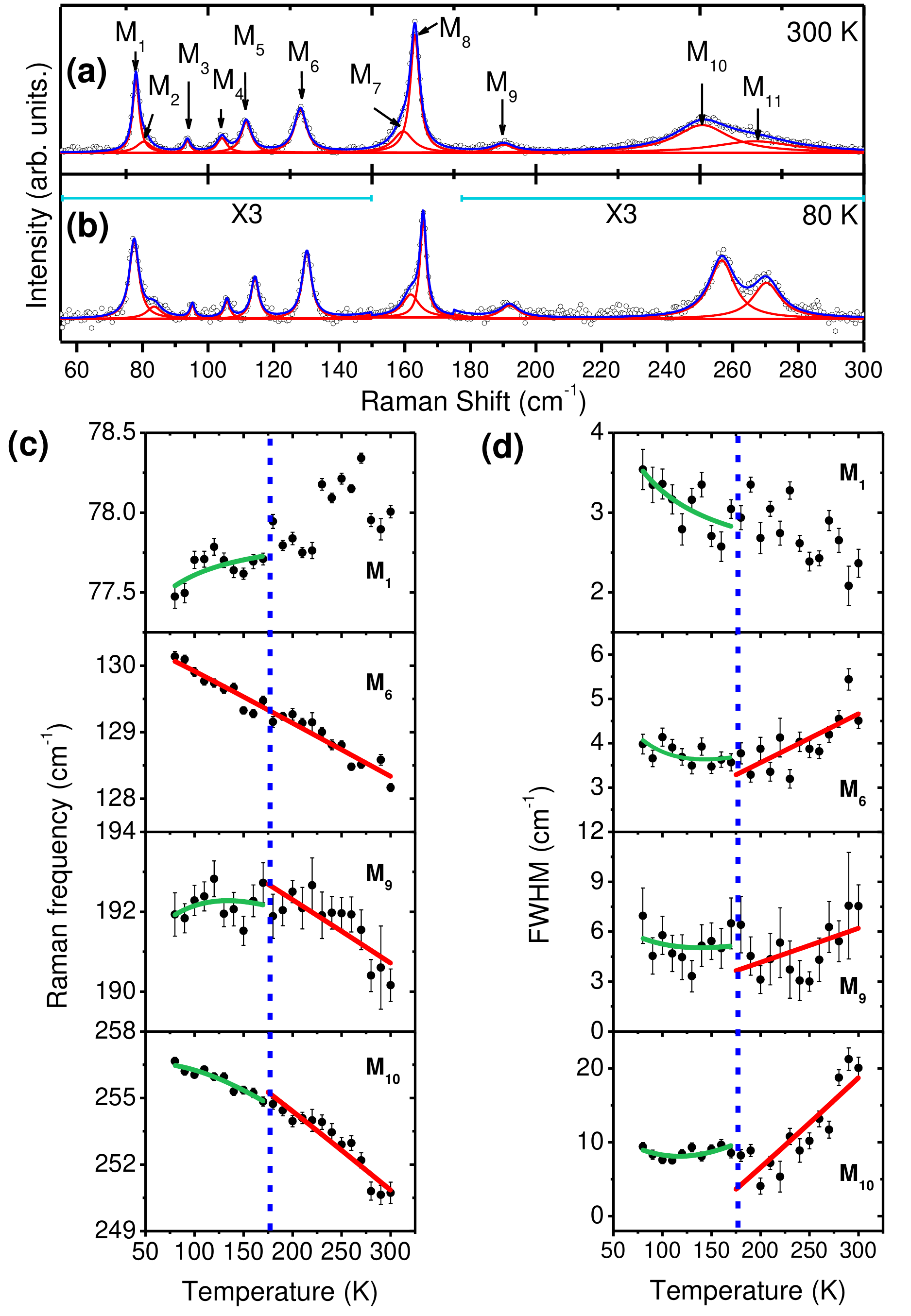}	
		\caption{ (Color Online) Raman spectra of 11 nm thin nanocrystal at (a) 300 and (b) 80 K. Black circles represents experimentally captured data points. Solid lines (Blue and red) are their respective Lorentzian fit. The peaks are shown from M$ _{1}$ to M$ _{11} $. Some regions at low temperature phase have been enhanced (ranges shown by green lines) for low signal response. (c) Peak position and (d) full width half maxima plot with temperature of M$ _{1} $, M$ _{6} $, M$ _{9} $ and M$ _{10} $ modes. The cubic anharmonic fitting is represented by the red line. Green lines represents the fitting from linear combination of cubic anharmonic and EPC contribution. Blue dashed line indicates the transition temperature ($T _{c} $) from 1T$ ^{\prime} $ to WSM phase.}
		\label{f1}	
	\end{figure*}
	\begin{figure}[b!]
		\includegraphics[width=1\columnwidth]{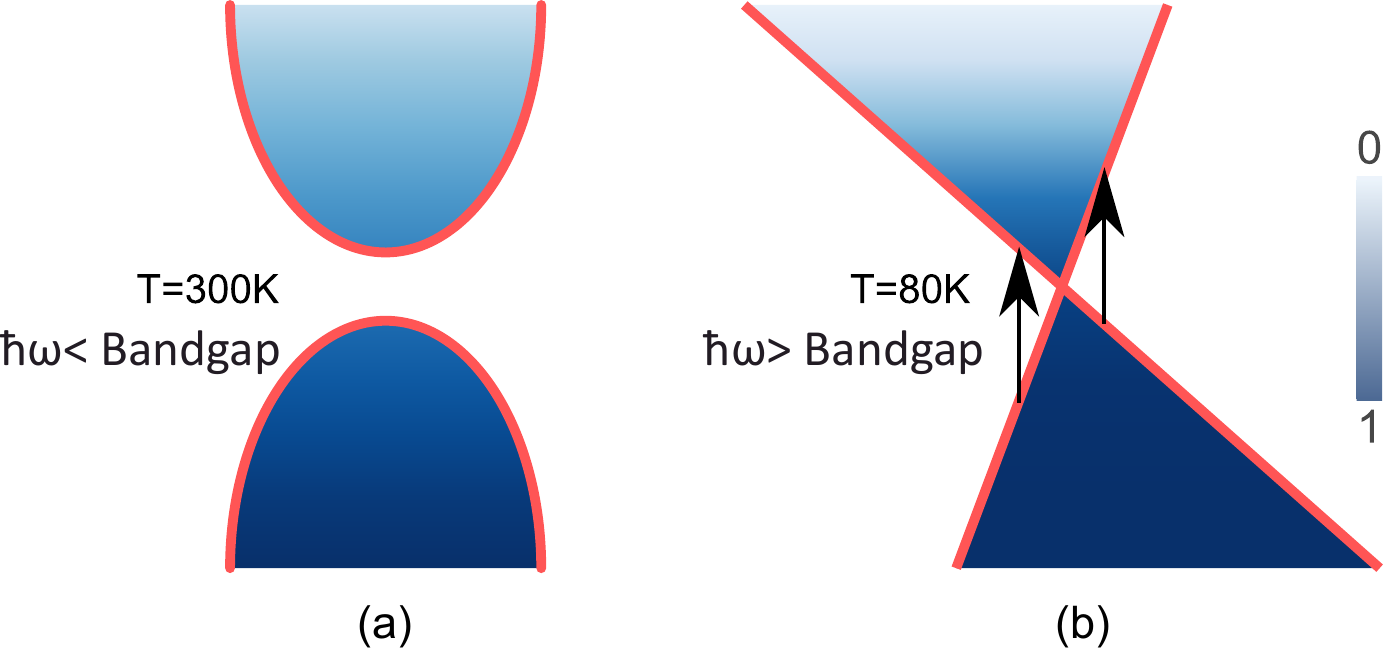}	
		\caption{ Schematic illustration of the electronic bands in (a) 1T$^{\prime} $ and (b) WSM phase of MoTe$ _{2} $. The arrows in the WSM phase represents electronic transition. The blue gradient represents the color map of the Fermi-Dirac distribution at 80K and 300K respectively.}
		
		\label{f2}
	\end{figure}
	
	At ambient  condition, 1T$^\prime$-MoTe$ _{2} $ belongs to P2$ _{1} $/m space group which gives 18 Raman active modes (12 A$ _{g}$+6 B$_{g} $) \cite{Brown:a04998}. Fig. \ref{f1} shows Raman spectra of 11 nm thick MoTe$ _{2} $ at 300 and 80 K, displaying 11 Raman modes, labeled as M$_{1}$ to M$ _{11} $. Spectra are fitted with a sum of Lorentzian functions to extract phonon frequency ($\omega$) and linewidth ($\gamma$) as a function of temperature, and plotted in Figs. \ref{f1}(c), \ref{f1}(d) and Fig. S3. Recent report shows appearance of an additional weak Raman mode near $ \sim $130 cm$ ^{-1} $ in the bulk T$ _{d} $ phase below the transition temperature 240 K \cite{zhang2016raman,PhysRevB.97.041410,chen2016activation}. However our measurements on 11nm thick 1T$ ^{\prime} $-MoTe$ _{2} $ do not show this additional new Raman modes down to 80 K. Theoretical and experimental results by Ma \textit{et al.} report that 1T$ ^{\prime} $ and T$ _{d} $ phases of MoTe$ _{2} $ have similar Raman spectra with two A$ _{g} $ modes near $ \sim $130 cm$ ^{-1}$ with a difference of only  $\sim$1 cm$ ^{-1} $ \cite{PhysRevB.94.214105}. It is very likely that this separation is not resolved in our 11nm thin crystal. At a first glance, the following observations are noteworthy in Figs. \ref{f1}(c) and (d): (i) The frequency and linewidth of the mode M$ _{1} $ show anomalous temperature dependence namely, the frequency decreases and linewidth increases with decreasing temperature. (ii) The linewidth of modes M$ _{6} $, M$ _{9} $ and M$ _{10} $ tend to show either saturation or increase below $T _{c} $. (iii) The frequency of modes M$ _{9} $ and M$ _{10} $ show marked deviation from linear dependence on temperature expected from cubic anharmonicity. To understand these trends, we recall that in a solid, a phonon can decay into two or more phonons (anharmonic contribution) or it can create a pair of electron and hole (electron -phonon coupling (EPC)) in conduction and valence bands respectively. These contribution to the phonon frequency and linewidth can be written as,
	\begin{equation}
		\omega(T)=\omega_{0}+\Delta\omega_{ph}+\Delta\omega_{e-ph}
		\label{w}
	\end{equation}
	\begin{equation}
		\gamma(T)=\gamma_0+\Delta\gamma_{ph}+\Delta\gamma_{e-ph}
		\label{g}
	\end{equation}
	
	In a simplified model, taking  cubic anharmonicity into account for the decay of a phonon into two phonons of equal energy, the temperature dependence is given by: $\Delta\omega_{ph}(T)=-A.g(T,\omega_{0})$ and $\Delta\gamma_{ph}(T)=B.g(T,\omega_{0})$, where $g(T,\omega_{0})=1+2/(e^{\hbar\omega_{0}/{2k_BT}}-1) $ with $ \omega_{0}$ being the phonon frequency at zero temperature, and $k_B$ is the Boltzmann constant \cite{PhysRevB.28.1928}, A and B are positive constants proportional to the phonon-phonon interaction strength. Neglecting the EPC contribution, the anharmonic decay model alone fails to capture the temperature dependence of $\omega$ for M$ _{9} $ and M$ _{10} $ and $\gamma$ for M$ _{6} $, M$ _{9} $ and M$ _{10} $ modes below $T _{c} $ (Figs. \ref{f1}(c) and (d)). In addition, M$ _{1} $ mode shows opposite temperature dependence of $\omega$ and $\gamma$ from cubic anharmonicity. We now argue that these anomalous changes below $T _{c} $ are due to strong EPC between Weyl Fermions and phonons in the WSM state.

	Fig. \ref{f2} shows a schematic of the electronic states near Fermi energy, above and below $T _{c} $.  The self-energy of the phonons due to EPC arises from the interband transition at the energy of the phonon mode ($ \hbar\omega_{0} $) as indicated by the arrows in Fig. \ref{f2}. At $T>T _{c} $, the semiconducting 1T$^\prime$ phase has a bandgap of $\sim$ 60 meV \cite{keum2015bandgap}. Since the bandgap is larger than the energies of the phonons, the electronic transitions at $ \hbar\omega_{0} $ do not occur and hence the contribution from the EPC to the phonon self-energy can be neglected. At $T<T _{c} $ in the gapless WSM  state (shown in Fig. \ref{f2}), the electron-hole excitations at the phonon energy $ \hbar\omega_{0} $ become feasible, resulting in considerable electron- phonon interaction. In  the WSM state, real part of the self-energy due to EPC contribute to the additional softening of the phonon frequency and the imaginary part to the additional linewidth of the Raman mode \cite{PhysRevB.29.2051}.
	Following Ref. \onlinecite{xu2017temperature},  the EPC contribution to phonon linewidth is, $\Delta\gamma_{e-ph}(T)=D.h(T,\omega_{0})$ where D is a positive constant and $h(T,\omega_{0})=[f(-\frac{\hbar\omega_{0}}{2})-f(\frac{\hbar\omega_{0}}{2})]$, with $f(x)$ representing the Fermi function. Assuming similar expression for the EPC contribution to the mode frequency, $\Delta\omega_{e-ph}(T)$ can be expressed as $-C.h(T,\omega_{0})$ with positive constant C. Eqs. (\ref{w}) and (\ref{g})  are fitted to the data (green line) below $T _{c} $ (Figs. \ref{f1}(c) and(d)). Table. S1 shows the extracted parameters with error bars. As seen in Fig. \ref{f1}, the inclusion of electron-phonon contribution to the modes along with cubic anharmonic contribution, captures the anomalous temperature dependence of $ \omega $ and $ \gamma $ below $T _{c} $, supporting the transition to WSM phase. However for the M$ _{1} $ mode, $\omega$ slightly increases and $\gamma$ slightly decreases (both by $\sim$ 0.5 cm$ ^{-1} $) even above $T _{c} $ and cannot be explained by our model and requires further study (Fig. \ref{f1}).     
	
	\section{Conclusion}
	In summary, we have carried out electrical transport measurements and Raman studies as a function of temperature in ultrathin 1T$^\prime$-MoTe$_2$. The nanocrystals with thickness larger than 11 nm show a resistance maximum which is attributed to a transition from 1T$^\prime$ to T$_d$ (WSM) state on cooling. Furthermore, the resistivity increase at low temperatures for $T<20 K$ shows $\rho\propto 1/T$, a signature of the WSM state. This is further supported by our $1/f$ electrical-noise studies. Temperature induced transition from topologically trivial 1T$^\prime$ phase to type-II Weyl semimetallic T$ _{d} $ phase is also captured by means of anomalous temperature dependence of few phonon frequencies and their linewidths due to contribution from the electron-phonon coupling in the WSM phase. We hope that our studies will motivate theoretical studies to quantitatively understand the absence of the WSM in ultrathin 1T$ ^{\prime} $-MoTe$ _{2} $ due to possible gap opening at the Weyl point and the temperature dependence of the $ 1/f $ power spectral density.

	\section*{Acknowledgement}
	The authors would like to acknowledge the Center for Nanoscience and Engineering (CENSE), IISc for fabrication facilities. AD and  AKS thank the Department of Science and Technology (DST) for funding with DSTO-2051 and Nanomission project as well as Year of Science Professorship.

	\bibliographystyle{apsrev4-2}
	\bibliography{ref}

\end{document}


\title{Supplemental Material}

\author{Manabendra Kuiri}
\thanks{These authors contributed equally}
\author{Subhadip Das}
\thanks{These authors contributed equally}
\author{D V S Muthu}
\author{Anindya Das}
\email{anindya@iisc.ac.in}
\author{A.K Sood}
\email{asood@iisc.ac.in}
\affiliation{Department of Physics, Indian Institute of Science, Bangalore 560012, India}

\maketitle
\section{Device Characterization}

\begin{figure*}[ht!]
	\includegraphics[width=0.9\textwidth]{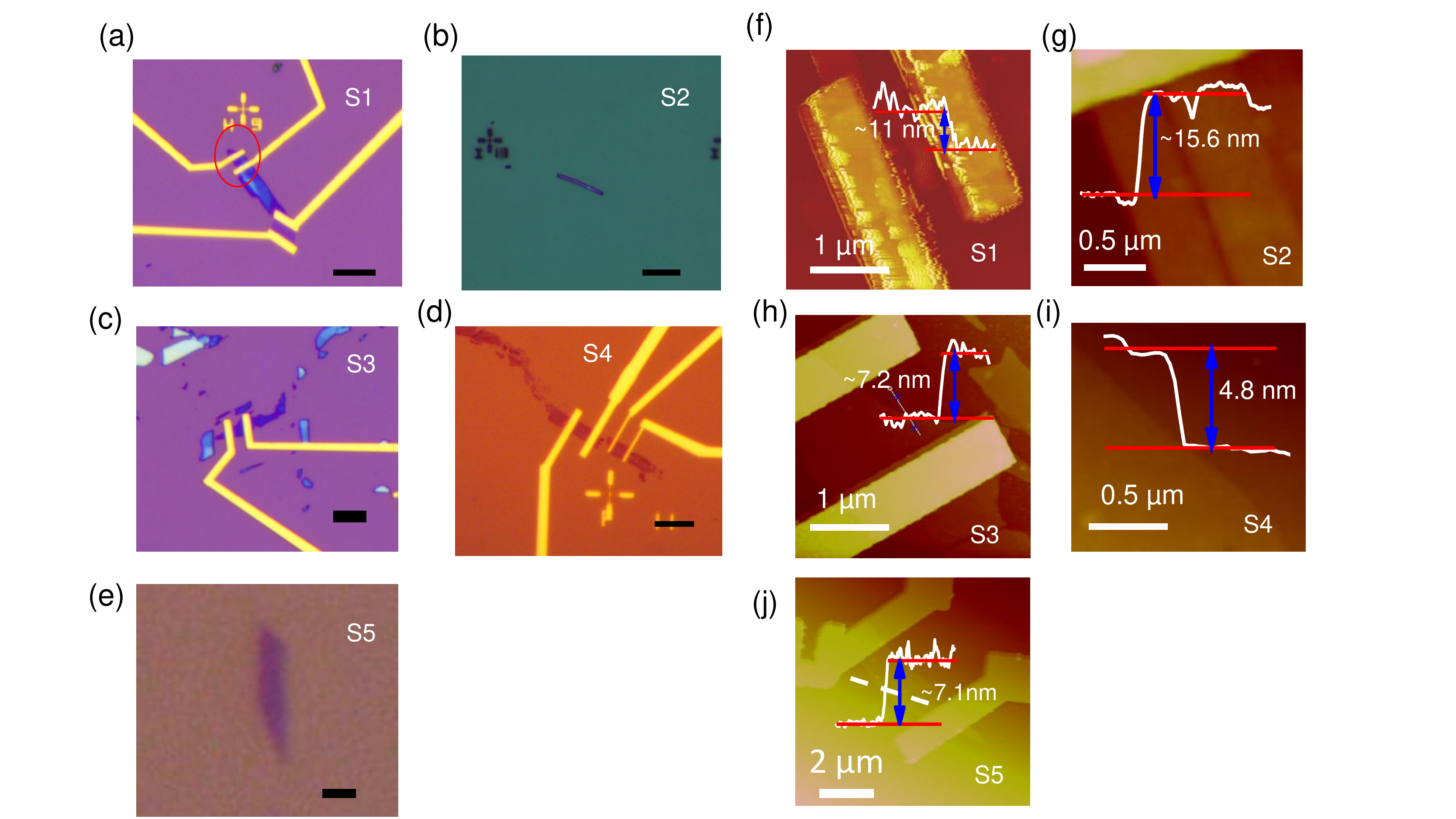}
	\caption{ Optical image of different 1T$ ^{\prime} $-MoTe$_2$ nanocrystals  with varying thickness (a) Sample S1, with thickness $\sim$11 nm. (b) Sample S2, with thickness $\sim$16 nm. (c) Sample S3, with thickness $\sim$7 nm, and (d) Sample S4, with thickness $\sim$5nm. (e) (d) Sample S5, with thickness $\sim$7nm. (f)-(i) AFM image of S1, S2, S3, S4 and S5. Inset shows their height profile.}
\end{figure*}

\newpage
\section{Additional transport Data}
\begin{figure*}[ht!]
	\includegraphics[width=0.8\textwidth]{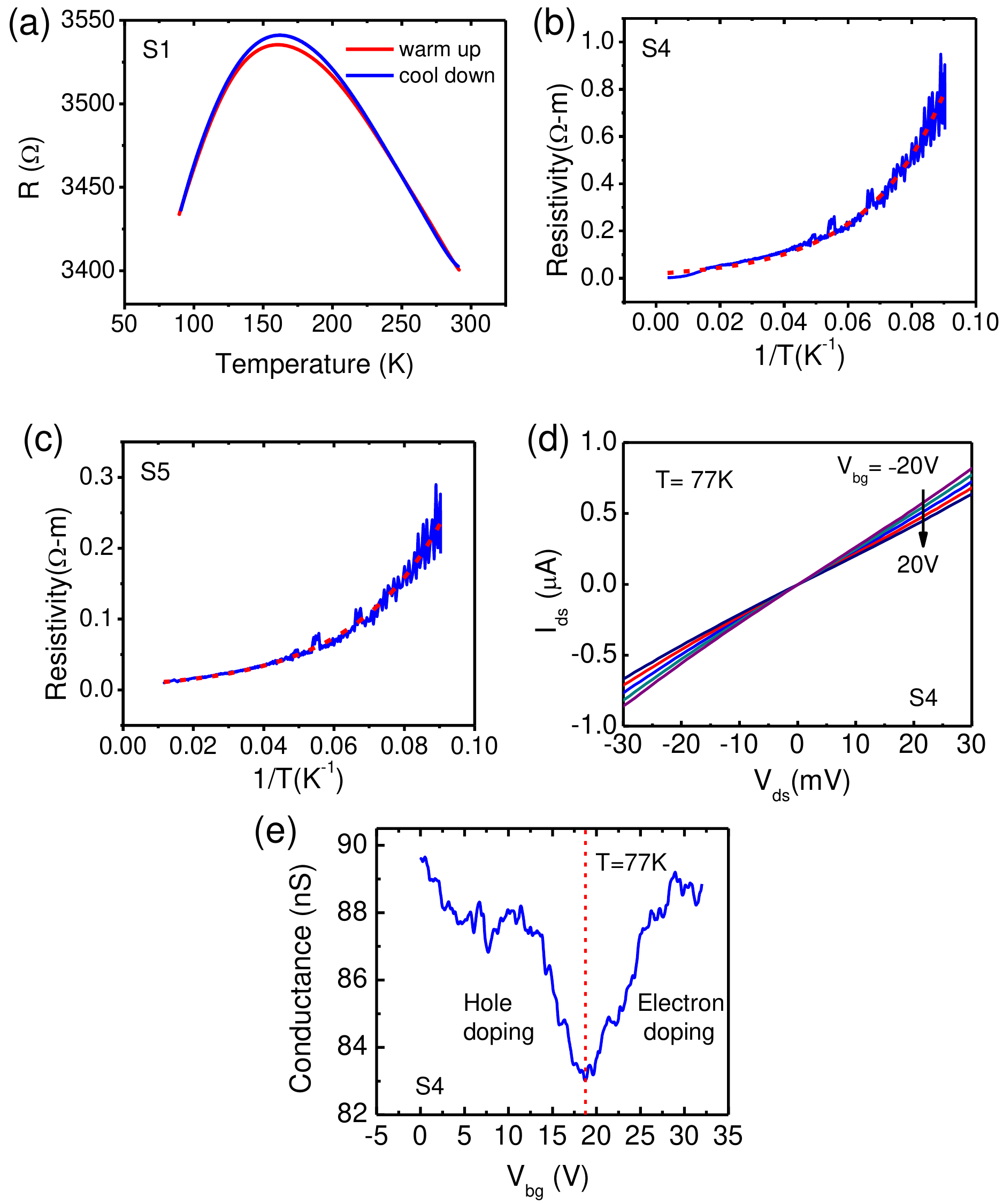}
	\caption{(a) Resistance versus temperature plot for warming up and cooling down for 11nm nanocrystal.  Resistivity versus temperature data (blue line) fit with Arrhenius equation (red dotted line) for (b) 5nm and (c) 7nm nanocrystals respectively. (d) Drain current (I$ _{ds} $) with drain voltage (V$ _{ds} $) plot for five different gate voltages of 5nm nanocrystal at 77K temperature. (e) Conductance of 5nm device plot with back gate voltage (V$ _{bg} $) at 77K. Dotted red line indicates V$ _{bg} $ for minimum conductance separating electron and hole dominated drain current.}
\end{figure*}

\newpage
\section{Additional Raman data}
\begin{figure*}[ht!]
	\includegraphics[width=0.7\textwidth]{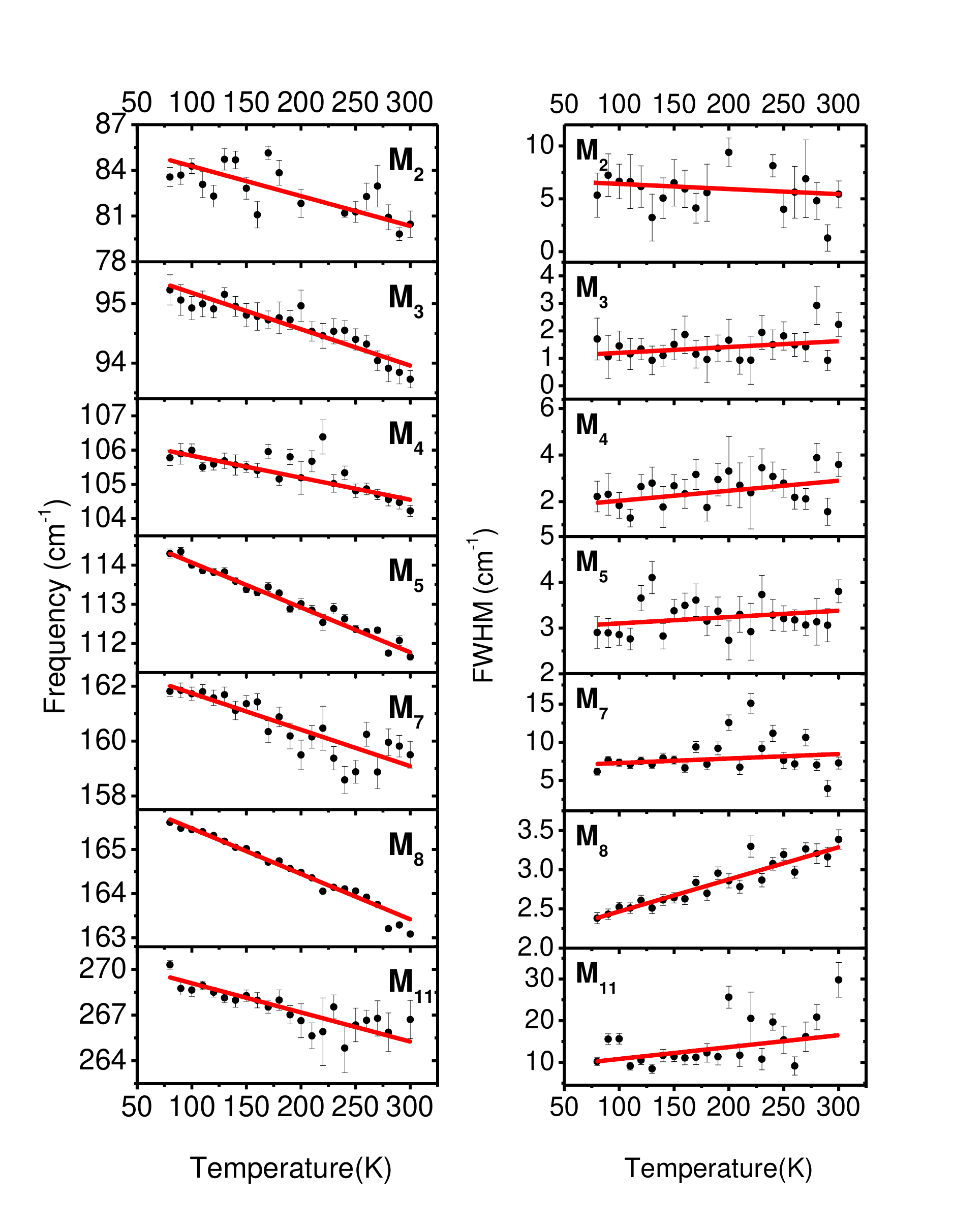}
	\caption{Variation of (a) Frequency and (b) FWHM with temperature of M$ _{2} $, M$ _3 $, M$ _{4} $, M$ _{5} $, M$ _{7} $, M$ _{8} $ and M$ _{11} $ mode. Red lines shows their respective cubic anharmonic fitting. }
\end{figure*}
\newpage
	
\begin{table}[h!]
	
	\begin{center}
		\caption{ Calculated value of the cubic anharmonic coefficients and electron-phonon coupling term}
		\begin{tabular}{|c|c|c|c|c|} 
			\hline
			Coefficients  &  M$ _{1} $ & M$ _{6}$ & M$ _{9}$ & M$ _{10}$  \\
			(cm$^{-1}$)  &&&& \\				 
			\hline
			A  &0 & 0.4$\pm$0.1 & 1.1$\pm$0.8 & 3.5$\pm$0.8 \\
			\hline
			B & 0 & 1.1 $\pm$0.7 & 1.5$\pm$0.6 & 12.0$\pm$4.5 \\
			\hline
			C &  1.07$\pm$0.4 & 0 & 5.1$\pm$1.7 & 3.5$\pm$1.3\\
			\hline
			D &  4.1$\pm$1.3 & 3.9$\pm$1.2 & 6.9$\pm$1.9 & 27.0$\pm$7.9\\
			\hline
		\end{tabular}
		
	\end{center}
\end{table}